\documentclass[aps,prl,twocolumn,floatfix]{revtex4-1}

\usepackage{graphicx}
\usepackage{amsmath}
\usepackage{upgreek}
\usepackage{mathtools}
\usepackage{braket}
\usepackage{bbold}

\begin{document}

\title{Observation of Rydberg-atom macrodimers: micrometer-sized diatomic molecules}

\author{Heiner Sa{\ss}mannshausen}
\author{Johannes Deiglmayr}
\affiliation{Laboratory of Physical Chemistry, ETH Zurich, Switzerland}
\email{jdeiglma@ethz.ch}

\newcommand{\AR}{{\textrm{A}}}
\newcommand{\BR}{{\textrm{B}}}

\begin{abstract}
Long-range metastable molecules consisting of two cesium atoms in high Rydberg states have been observed in an ultracold gas. A sequential three-photon two-color photoassociation scheme was employed to form these molecules in states, which correlate to $np(n+1)s$ dissociation asymptotes. Spectral signatures of bound molecular states are clearly resolved at the positions of avoided crossings between long-range van der Waals potential curves. The experimental results are in agreement with simulations based on a detailed model of the long-range multipole-multipole interactions of Rydberg-atom pair states. We show that a full model is required to accurately predict the occurrence of bound Rydberg macrodimers. The macrodimers are distinguished from repulsive molecular states by their behavior with respect to spontaneous ionization and possible decay channels are discussed.
\end{abstract}

\maketitle

A new regime of ultracold chemistry has been established by the possibility to routinely produce ultracold atomic samples with translational temperatures below 1~mK using laser cooling\,\cite{hansch1975,wineland1979} and trapping\,\cite{raab1987,grimm2000}. The characteristic feature of this regime is that reactions are not driven by thermodynamical quantities, \textit{i.e.} temperature or pressure, but by the precise manipulation of the internal quantum states of the constituents and the interactions between them. Landmark results in the emerging field of ultracold chemistry include the formation of ultracold molecules in the absolute ground state using photoassociation~\cite{fioretti1998,viteau2008,deiglmayr2008} or magneto-association~\cite{ni2008,danzl2010}, the control of chemical reactions by manipulating the quantum statistics of the reactants~\cite{ospelkaus2010,knoop2010}, or the photodissociation of molecules with full control over reactant and product channels~\cite{mcDonald2016}. The enabling factor in all these studies is the control of long-range interaction potentials, which dominate the dynamics at low collision energies.

\begin{figure}
\begin{center}
\includegraphics[width=0.99\linewidth]{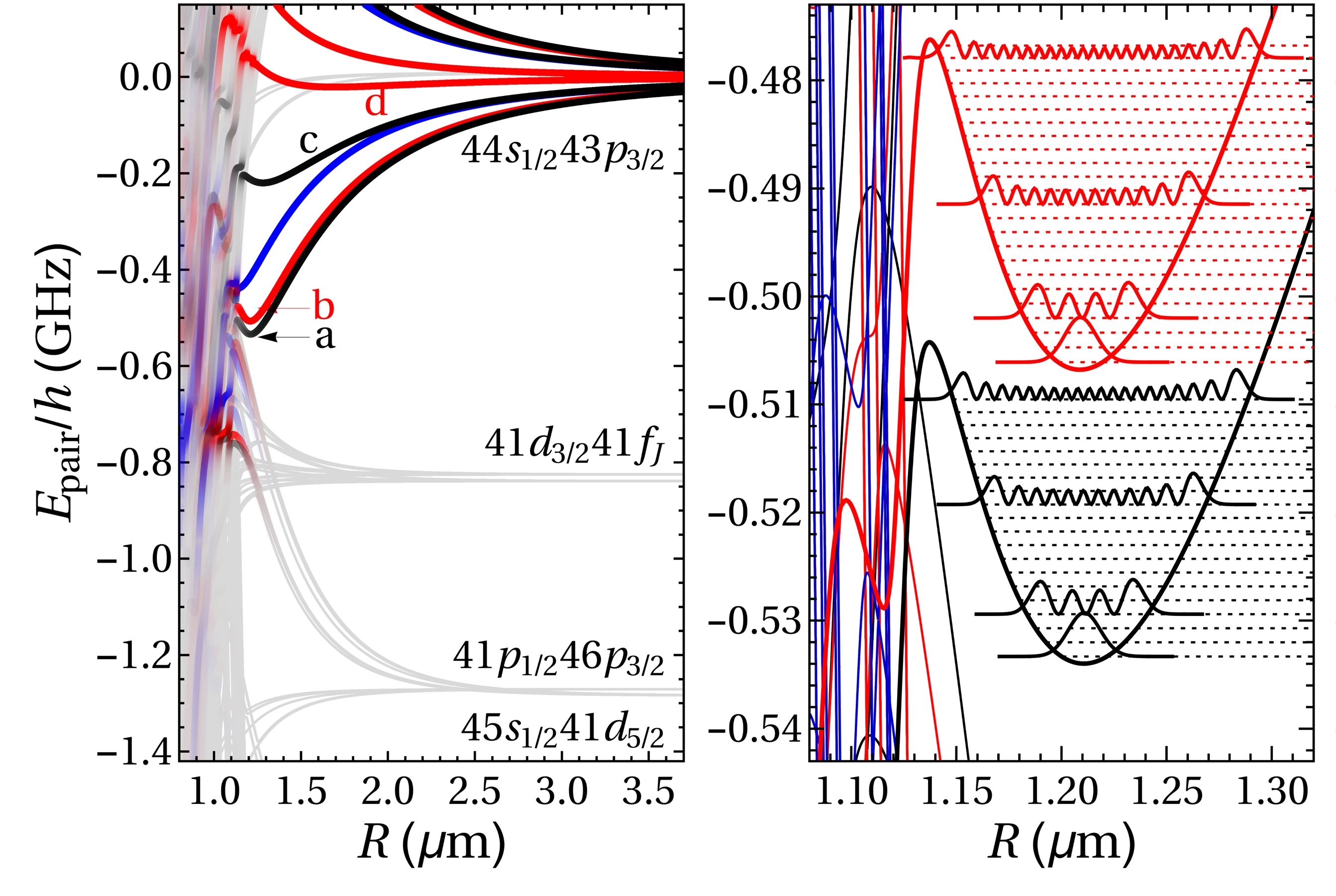}
\caption{\label{fig:pecs}(Color online) Left panel: Potential-energy curves of two Cs Rydberg atoms in the vicinity of the $43p_{3/2}44s_{1/2}$ dissociation asymptote. Black, red, and blue curves correspond to the molecular states with $\Omega$=0,1, and 2, respectively. The intensity of the color codifies the value of the $43p_{3/2}44s_{1/2}$ character (gray corresponding to zero, full color to more than 0.5~\% overlap with the asymptotic pair state). Bound macrodimer states are labeled by letters a to d. Right panel: the states a (lower curve) and b (upper curve) marked in the left panel are shown on a magnified scale. Energies (dashed lines) and probability densities of selected vibrational eigenstates are indicated.}
\end{center}
\end{figure}

An extreme case of long-range interactions are van der Waals forces between two atoms in Rydberg states, scaling in the case of an off-resonant dipole-dipole interaction with the internuclear separation $R$ as $R^{-6}$ and with the principal quantum number $n$ as $n^{11}$~\cite{gallagher2008}. Metastable molecular states with internuclear separations exceeding 1~$\upmu$m, so called macrodimers, are predicted to result from these interactions~\cite{boisseau2002, schwettmann2007, kiffner2012}. The molecular potential minima supporting these bound states arise from avoided crossings between long-range potential-energy curves correlating to different dissociation asymptotes of the doubly-excited dimers (see Fig.~\ref{fig:pecs}). Astonishingly, such macrodimers are predicted to have lifetimes limited by radiative decay of the constituent atomic Rydberg states~\cite{marcassa2014}, even though they are energetically located in the Cs\,+\,Cs$^+$\,+\,e$^{-}$ ionization continuum and the density of electronic states is extremely high, two conditions one would expect to lead to fast autoionization~\cite{lefebvre2004}. Experimentally, these molecular states have remained elusive. Molecular resonances were observed in the Rydberg-excitation spectrum of ultracold gases and were identified as arising from the correlated excitation of two interacting Rydberg atoms\,\cite{farooqi2003,overstreet2009,deiglmayr2014,sassmannshausen2015}. However, evidence for bound Rydberg-atom macrodimers could only be obtained indirectly from dynamics following delayed pulsed-field ionization (PFI) for molecular states in a particular electric field\,\cite{overstreet2009}, and from the dynamics during the Penning ionization of excited atom pairs\,\cite{sassmannshausen2015}. In this Letter, we present the observation of Rydberg-atom macrodimers by the unambiguous assignment of photoassociation resonances to long-range potential minima, calculated on the basis of a detailed model of the long-range interactions\,\cite{sassmannshausen2015}. An investigation of the lifetime of the macrodimers and a discussion of possible decay channels support the assignment.

The investigated macrodimer states were first identified in an extensive theoretical search for minima in the long-range potential functions of Cs Rydberg-atom pair states, correlating to atomic $ns$, $np$, and $nd$ Rydberg states. The employed model is based on the multipole expansion of the static Coulomb interaction (truncated after terms corresponding to octupole-octupole interactions) and a basis constructed of atomic orbitals~\cite{sassmannshausen2015}, where the only parameters are the atomic quantum defects taken from~\cite{goy1982}. Contributions from electronic-parity-violating terms such as the dipole-quadrupole interaction, enabled by a coupling of quasi-degenerate rotational levels, are included in an effective manner~\cite{deiglmayr2014}. The only remaining molecular symmetry is associated with the projection of the total electronic angular momentum on the internuclear axis with quantum number $\Omega$. The model is valid for interatomic distances larger than the LeRoy radius $R_\textrm{LR}=2\left(\braket{r_\AR^2}^{1/2}+\braket{r_\BR^2}^{1/2}\right)$\,\cite{leroy1973}, which is <~0.6~$\upmu$m for the states considered in this Letter. The identification of possible candidates for an experimental investigation was based on the following criteria: \textit{i}) The equilibrium internuclear distance of the bound state exceeds 1\,$\upmu$m, the most likely next-neighbor distance at a ground-state-atom density of $10^{11}$~cm$^{-3}$, \textit{ii}) the vibrational frequency exceeds 1~MHz which allows to spectrally resolve vibrational levels, and \textit{iii}) the spectral signatures of the bound states are clearly isolated from other resonances. The most promising candidates for the observation of macrodimer states in Cs  under these restrictions are states correlated to $np_{3/2}(n+1)s_{1/2}$ pair-state asymptotes, with $n=43, 44$ (see Fig.~\ref{fig:pecs}). An important finding of this theoretical study is that general predictions for the existence of macrodimer states by more simplified models (\textit{e.g.} Ref.~\cite{kiffner2014} applied to Cs atoms in zero field and Ref.~\cite{samboy2011} for Rb) could not be confirmed in our extended model.

\begin{figure}
\begin{center}
\includegraphics[width=0.99\linewidth]{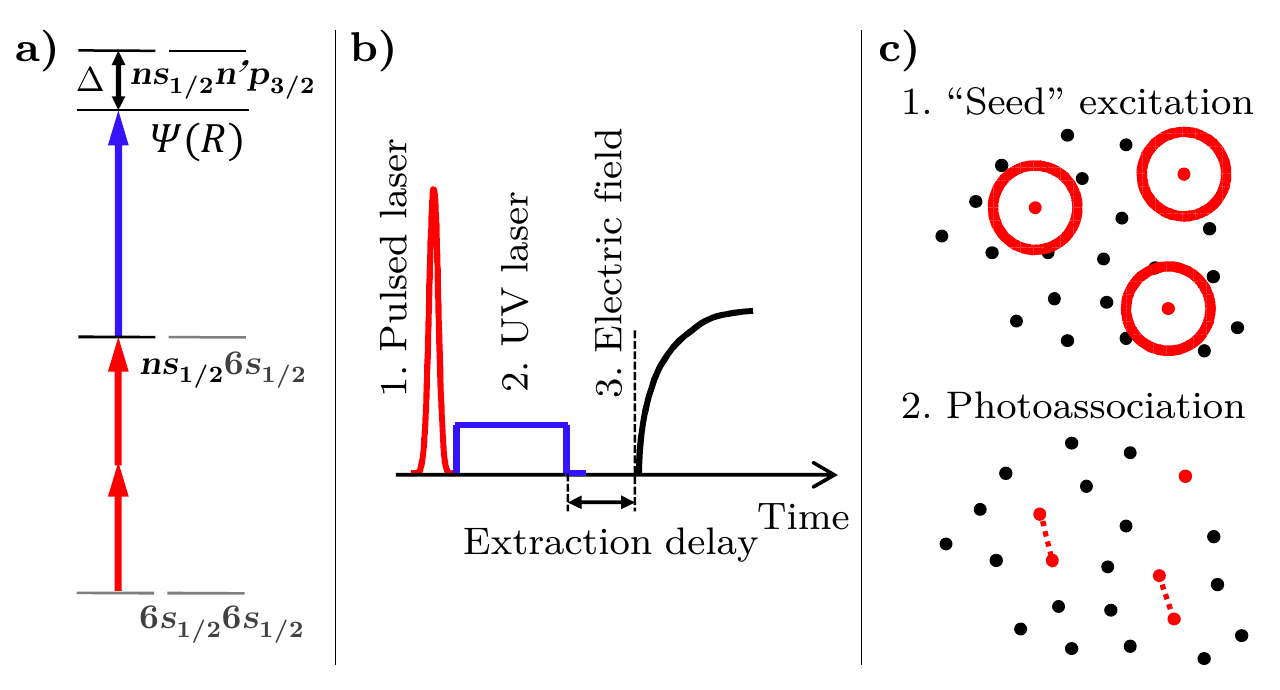}
\caption{\label{fig:scheme} (Color online) a) Energy-level diagram for the two-color three-photon photoassociation of ground-state atoms to a doubly-excited molecular Rydberg state $\Psi(R)$, b) sequence of laser excitation and detection by pulsed-field ionization (PFI), c) sketch of the two-step excitation process, where black and red dots represent ground-state and Rydberg atoms, respectively, and the red circle illustrates the extent of the vibrational wavefunction of the macrodimer.}
\end{center}
\end{figure}

The experimental scheme for the formation of macrodimers by photoassociation is based on a sequential excitation of the doubly excited molecular state (see Fig.\,\ref{fig:scheme})~\cite{sassmannshausen2016}. In a first step, a small number of atoms is resonantly excited into $(n+1)s_{1/2}$ Rydberg states. These atoms, which are referred to as ``seed'' Rydberg atoms in the following, modify the Rydberg-excitation spectrum of ground-state atoms in their vicinity. The resulting interaction-induced shifts $\Delta$ (see Fig.~\ref{fig:scheme}a)) suppress the excitation of further atoms within a certain volume around the seed atom and are the basis of the so-called dipole blockade mechanism~\cite{lukin2001}. The accessible interaction potentials (Fig.~\ref{fig:pecs}), averaged over the distribution of internuclear separations, can be probed by applying a second, detuned laser pulse and monitoring the additional excitation of ground-state atoms~\cite{reinhard2008}. To form macrodimers, the frequency of the second excitation laser is tuned into resonance with the photoassociation transition to the molecular state, which is detuned by $\Delta$ from the transition to the isolated $np_{3/2}$ Rydberg state. A seed and a ground-state atom can then be photoassociated into a macrodimer when their scattering wavefunction overlaps with the excited molecular wavefunction (Fig.\,\ref{fig:scheme}c))~\cite{thorsheim1987}. It is noteworthy that this process is a true two-body transition in which the state of both atoms is changed, similar as in laser-induced collisional energy transfers (LICET)~\cite{bambini1994}.

The experiments are performed with dense ($1\cdot10^{12}$~cm$^{-3}$), ultracold ($T \approx 40$~$\upmu$K) samples of Cs atoms, released from a crossed optical dipole trap. Only atoms in the densest part of the sample are selected for Rydberg excitation by optical pumping into the $F=4$ hyperfine component of the ground state. Details of the experimental setup can be found in Refs.~\cite{sasmannshausen2013,sassmannshausen2015}. The $ns_{1/2} \leftarrow 6s_{1/2}$ two-photon transition of the excitation scheme outlined in Fig.\,\ref{fig:scheme} is driven by the pulse-amplified output of a diode laser (Toptica DL-100 pro), yielding pulses with a duration of 7~ns and a near-transform-limited bandwidth of 60~MHz at wavelengths around 639~nm. The photoassociation transition is driven by the continuous-wave frequency-doubled output of a ring dye laser (Coherent 899 and MDB-200) with a wavelength of 319~nm and a bandwidth of 1.5~MHz, from which UV laser pulses of typically 1-10~$\upmu$s duration and peak powers exceeding $200$~mW are obtained using an acousto-optical modulator. The frequencies of both lasers are stabilized to a commercial wavemeter (HighFinesse WS-7) referenced to a frequency comb (Menlo Systems  FC1500-250-WG)~\cite{deiglmayr2016}. Both laser beams are focused to spot sizes of $\sim 150$\,$\upmu$m in diameter and overlapped spatially at right angles in the Cs atom cloud (there is no temporal overlap between the two laser pulses). Transitions are detected by extracting Cs$^+$ ions formed spontaneously or by PFI of Rydberg atoms or molecules with a rising electric field (rise time $\sim$ 1~$\upmu$s), and monitoring the resulting ion yield using a microchannel-plate detector. Spontaneously-formed ions and field-ionized Rydberg atoms in different quantum states are discriminated by their time of flight to the detector.  Stray electric and magnetic fields are compensated to below 25\,mV/cm and 20\,MG, respectively, where their influence on the Rydberg energy spectrum is negligible for the relevant states.

\begin{figure}
\begin{center}
\includegraphics[width=0.99\linewidth]{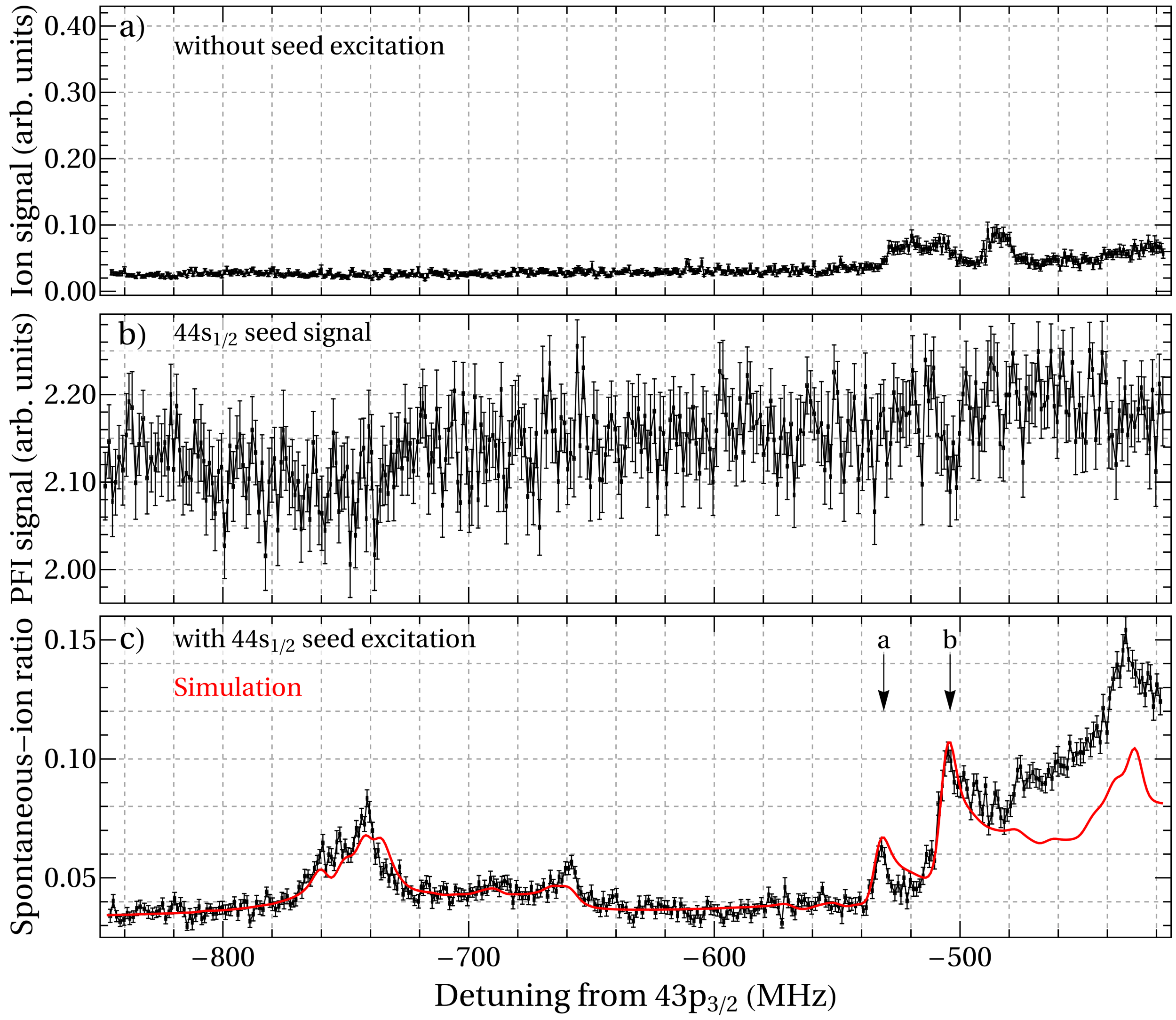}
\caption{\label{fig:43p44s} (Color online) Ion signals (black dots) as a function of the detuning from the transition $43p_{3/2} \leftarrow 6s_{1/2}$: (a) Spontaneous-ion signal when no seed atoms are present, (b) PFI signal when $44s_{1/2}$ seed Rydberg atoms are present during excitation, and (c) ratio of spontaneous-ion to PFI signals when $44s_{1/2}$ seed Rydberg atoms are present (the background signal shown in panel a) was subtracted). The red line shows a simulated spectrum for a linewidth of 7~MHz full width at half maximum, as discussed in the text. Labels a and b indicate the frequencies of transitions to the minima of the macrodimer potentials marked by a and b in Fig.~\ref{fig:pecs}.}
\end{center}
\end{figure}

Experimental spectra recorded below the $43p_{3/2}44s_{1/2}$ dissociation asymptote are presented in Fig.\,\ref{fig:43p44s}. Fig.\,\ref{fig:43p44s}a) shows the spontaneous-ion signal, extracted after a delay of 5~$\upmu$s, as a function of the detuning of the UV laser from the atomic  $43p_{3/2} \leftarrow 6s_{1/2}$ transition with no seed excitations present (UV pulse duration 5~$\upmu$s). A set of broadened resonances at detunings around $-500$\,MHz is attributed to the excitation of long-range Rydberg molecules bound by electron-atom scattering~\footnote{This assignment is based on the observation that molecular Cs$^+_2$ ions are detected on resonance, which we empirically find characteristic of Cs$_2$ long-range Rydberg molecules bound by electron-atom scattering~\cite{Greene2000,Bendkowsky2009}.}. The increasing signal for higher frequencies is attributed to single-color two-photon excitation of Rydberg-atom pairs. The spectra shown in Fig.\,\ref{fig:43p44s}b) and c) were recorded when about 50 seed excitations (corresponding to a density of less than $5\cdot 10^{7}$~cm$^{-3}$) were present during the UV laser pulse. The ion signal resulting from PFI of $44s_{1/2}$ seed Rydberg atoms is presented in Fig.\,\ref{fig:43p44s}b). The fluctuations of this signal (on the order of 5 ions after averaging over 30 laser-excitation and detection cycles) are caused by intensity instabilities of the pulse-amplified laser and mask additional excitations by the UV laser. As in previous works, we observe that interacting Rydberg-atom pair states in the relevant range of detunings and distances ionize within a few $\upmu$s, leading to the appearance of Cs$^+$ ions~\cite{deiglmayr2014,sassmannshausen2015}. Surprisingly, the photoassociated macrodimers are also found to ionize within 5~$\upmu$s. The ratio of spontaneous-ion to PFI signal shown in Fig.\,\ref{fig:43p44s}c) is therefore taken as a measure of the number of additional excitations per seed atom.
In comparison to the background measurement (Fig.\,\ref{fig:43p44s}a)), several additional resonances are present.

In order to interpret the experimental observations, we simulate the spectrum on the basis of the calculated potential-energy curves (see Fig.~\ref{fig:pecs})~\cite{deiglmayr2014,sassmannshausen2015}. Assuming a resonant single-photon photoassociation, the expected additional signal at laser frequency $\nu = E/h$ is proportional to
\begin{equation}
\label{eq:lineprofile}
    s(E) = \sum_\Psi \int_0^\infty \omega_\textrm{atom}^2 G(E-E_\Psi(R))\overline{p_\Psi}(R) R^2 \textrm{d}R,
\end{equation}
where $\omega_{\rm atom}$ is the Rabi frequency for the atomic $np_{3/2} \leftarrow 6s_{1/2}$ transition, $G(E)$ is the laser line profile, for which a Gaussian function was chosen, and $\overline{p_\Psi}(R)$ is the sum of the squared $np_{3/2}(n+1)s_{1/2}$ coefficients of the molecular state $\Psi$ with energy $E_\Psi(R)$. This model is valid when the mean number of additional excitations per seed atom is much smaller than 1, and the relative motion of the atoms can be treated classically. The latter assumption holds because the extent of the vibrational wavefunctions of the macrodimers (see Fig.~\ref{fig:pecs}, right panel) is large compared to the thermal de Broglie wavelength of the atoms ($\lambda_{\mathrm{th}}\sim30$~nm). The simulated spectrum, shown as a red curve in Fig.\,\ref{fig:43p44s}c), reproduces all features of the experimental spectrum and allows for an unambiguous assignment of the observed resonances to molecular states. The resonances at detunings of $-530(3)$~MHz and $-507(3)$~MHz arise from the potential minima of the macrodimer states with $\Omega=0$ and 1, labeled by letters a and b, respectively, in Figs.~\ref{fig:pecs} and \ref{fig:43p44s}. The broad resonance centered at $-430$~MHz results from avoided crossings of molecular states with $\Omega=2$. The resonances below $-600$\,MHz arise from the maxima of potential-energy curves that result from the same avoided crossings leading to the existence of the macrodimer states. For detunings above $-480$~MHz, the model underestimates the fraction of additional excitations. We attribute this deviation to increasing contributions from an excitation-avalanche mechanism as discussed in Refs.~\cite{garttner2013,lesanovsky2014,schempp2014,malossi2014,urvoy2015}. The experimental spectra are reproduced best by a line profile $G(E)$ with a width of 7\,MHz, which is much larger than the measured linewidth of 2~MHz for an atomic transition. The additional broadening is very likely caused by the thermal distribution of relative velocities in the atomic sample~\cite{jones1999} (contributing about 1~MHz) and by the interactions with all other seed atoms. The latter contribution is estimated to be 3~MHz from the measured broadening of the $43p_{3/2} \leftarrow 6s_{1/2}$ transition in the presence of seed atoms with similar densities as in the measurements presented in Fig.~\ref{fig:43p44s}. These broadenings mask the discrete resonance structures for photoassociation into vibrational levels with a spacing of 1.3~MHz (see Fig.~\ref{fig:pecs}).

\begin{figure}
\begin{center}
\includegraphics[width=0.99\linewidth]{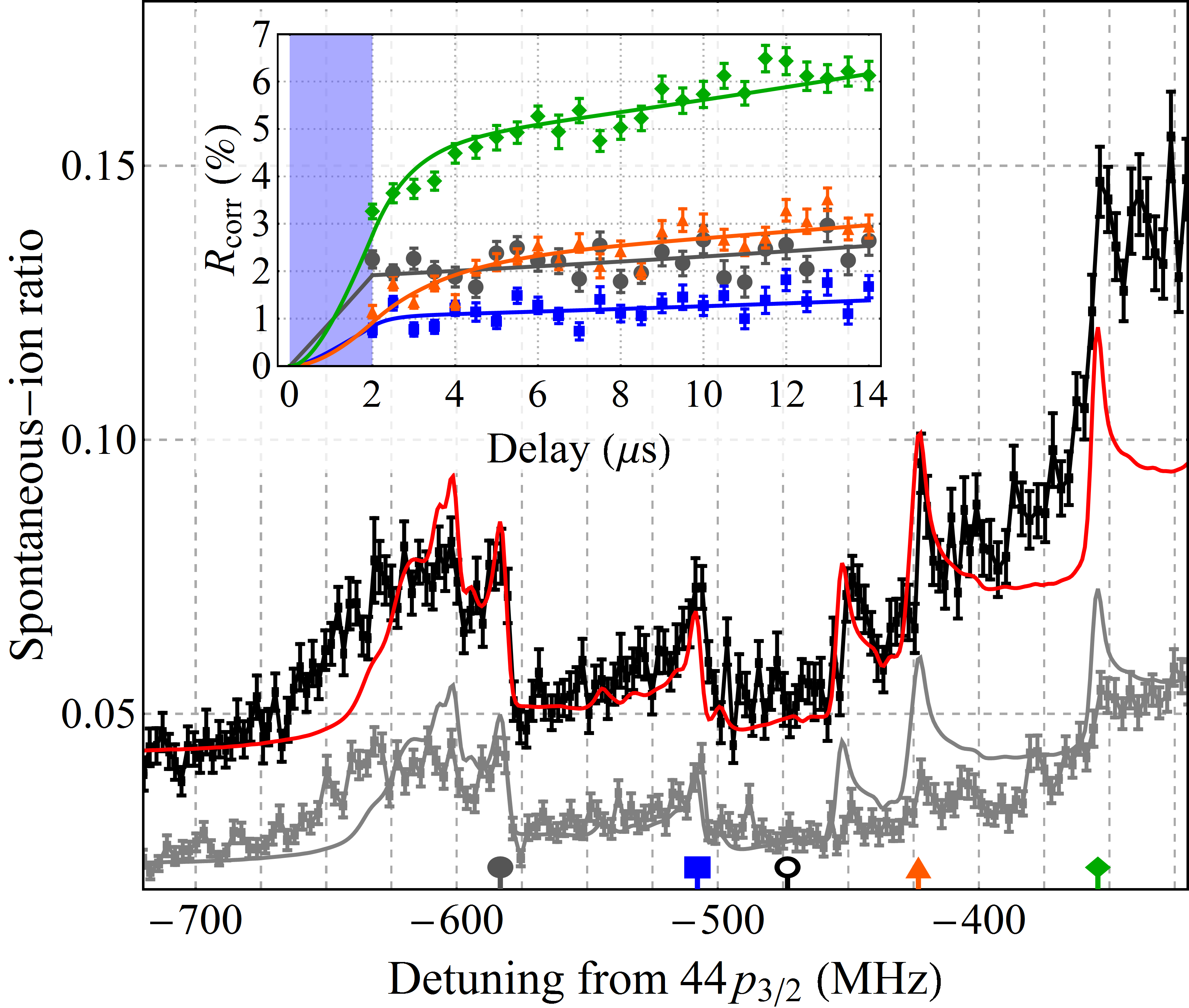}
\caption{\label{fig:44p45s} (Color online) Ratio of spontaneous-ion to PFI signals as function of the detuning of the UV laser (2\,$\upmu$s pulse length) from the $44p_{3/2} \leftarrow 6s_{1/2}$ transition when $45s_{1/2}$ seed atoms are present. Ions were extracted after a delay of 0 (gray points) and 12~$\upmu$s (black points). The spectrum simulated for a linewidth of 5~MHz (vertically scaled and offset) is shown as solid red and gray lines for comparison. The inset shows the measured ratios (circles, squares, triangles, and diamonds) as a function of the extraction delay at the fixed laser detunings marked by the corresponding symbol on the frequency axis. A time-dependent background signal (measured at the frequency indicated by an open circle) was subtracted from all curves. Solid curves show the fit of a rate-equation model to the data, the light-blue shaded region indicates the UV-excitation pulse.}
\end{center}
\end{figure}

 Photoassociation resonances of bound macrodimers are also observed on the low-frequency side of the $44p_{3/2}45s_{1/2}$ dissociation asymptote, as shown in Fig.\,\ref{fig:44p45s}. The dependence of the spontaneous-ion ratio on the extraction delay reveals clear differences between resonances arising from the photoassociation of macrodimers, and from the excitation of Rydberg-atom pairs in the continuum. The spectra in Fig.\,\ref{fig:44p45s} were recorded after excitation of about 50 atoms to the $45s_{1/2}$ state, followed by a 2-$\upmu$s-long UV pulse. The spectrum presented in black was obtained by extracting ions after a delay of 12\,$\upmu$s following UV excitation, whereas no delay was introduced in the measurement depicted in gray. The spectra exhibit an overall, UV-frequency-independent offset that originates from spontaneous ionization of seed atoms. While the exact ionization mechanism is unclear, the observed rate of 1.4~kHz is compatible with calculations for black-body-enhanced field ionization of $ns_{1/2}$ Cs Rydberg states~\cite{beterov2009}. More interestingly, the resonances originating from bound states at detunings of $-450$, $-425$, and $-355$~MHz are clearly visible after an extraction delay and are not present, or much weaker, in the spectrum recorded without delay. Resonances originating from potential maxima, \textit{e.g.} at $-583$ and $-507$~MHz, are observed with similar strengths under both measurement conditions. These observations indicate that the bound macrodimers decay more slowly into Cs$^+$ ions than Rydberg-atom pairs excited to continuum states. The inset of Fig.\,\ref{fig:44p45s} depicts measurements of the background-corrected spontaneous-ion ratios $R_\textrm{corr}$ as a function of the extraction delay for different UV-laser detunings together with results of a rate-equation model for the decay and formation processes of seed atoms and macrodimers. The free parameters of this model, the photoassociation and ionization rates at a given detuning, are determined from a fit of the model to the data. The model reveals that the apparent increase of $R_\textrm{corr}$ observed for all detunings results from the normalization to the PFI signal, which decreases with time due to radiative decay (and to a much lesser extent due to ionization) of the initial seed atoms. The fitted ionization rates exceed 3~MHz at the positions of the resonances located at detunings of $-583$\,MHz and $-507$\,MHz. Significantly lower ionization rates of about 0.6(1) and 1.0(1)~MHz are obtained for the resonances at detunings of $-421$ and $-355$\,MHz, respectively, which correspond to bound macrodimers. Although the corresponding lifetimes of less than 2~$\upmu$s are much shorter than expected~\cite{boisseau2002,marcassa2014}, they are still considerably longer than at the positions of potential maxima, which we consider a further indication of the bound nature of these states.

 Whereas the decay of Rydberg-atom pairs excited to attractive potentials can be understood in a Penning-ionization model~\cite{viteau2008b,sassmannshausen2015}, the calculated potentials presented in Fig.\,\ref{fig:pecs} indicate that such a process cannot occur for the bound macrodimers. Other conceivable ionization mechanisms are \textit{i}) electronic transitions related to inter-Coulombic decay, in which one atom is transferred into the (ionization) continuum while the other atom is de-excited into a more strongly bound state~\cite{cederbaum1997,amthor2009}, and \textit{ii}) vibrational autoionization, where the decay of the bound macrodimer is caused by non-adiabatic couplings to molecular states with open ionization channels~\cite{lefebvre2004}. By extending the models for mechanism \textit{i}) and \textit{ii}) described in Ref.~\cite{amthor2009} and Ref.~\cite{samboy2011}, respectively, to higher-order interaction terms and larger basis sets, we verified that neither mechanism can explain decay rates exceeding 1~kHz for the macrodimer states investigated here, even for vibrationally excited levels. The rates for \textit{ii}) are, however, expected to increase significantly when interactions with a third atom, breaking the symmetry associated with $\Omega$ and further increasing the density of accessible rovibronic states, become important. The increased linewidth of the molecular resonances discussed above indicates that the interaction with other seed atoms might be the reason for the observed fast decay rates.

 In summary, we find that while the treatment of Rydberg-Rydberg interactions on the basis of the multipole-expansion of the static Coulomb interaction yields very accurate adiabatic potential-energy functions, the dynamical evolution of the formed macrodimers clearly reveals the importance of non-adiabatic couplings and few-body interactions for strongly correlated Rydberg systems. Future studies with single atoms in optical dipole traps with tunable separations~\cite{beguin2013,hankin2014}, or atoms in an optical lattice~\cite{schaus2012}, should enable the investigation of the importance of few-body interactions and probe the isolated dimer with the prospect of resolving the vibrational structure of macrodimers.

\begin{acknowledgments}
 This work is supported financially by the ETH Research Grant ETH-22 15-1, the Swiss National Science Foundation under Project Nr.~200020-159848, the NCCR QSIT of the Swiss National Science Foundation, and the EU Initial Training Network COHERENCE under grant FP7-PEOPLE-2010-ITN-265031. We acknowledge the European Union H2020 FET Proactive project RySQ (grant N. 640378). We thank Prof. Fr\'ed\'eric Merkt (ETH Zurich) for critical reading of the manuscript and for continuous and exemplary support. The authors acknowledge fruitful discussions with Prof. Pierre Pillet (Laboratoire Aimé Cotton, Orsay) and JD acknowledges instructive discussions with PD Alexander Kuleff and Dr. Kirill Gokhberg (both Institute of Physical Chemistry, University of Heidelberg).
\end{acknowledgments}

\begin{thebibliography}{49}%
\makeatletter
\providecommand \@ifxundefined [1]{%
 \@ifx{#1\undefined}
}%
\providecommand \@ifnum [1]{%
 \ifnum #1\expandafter \@firstoftwo
 \else \expandafter \@secondoftwo
 \fi
}%
\providecommand \@ifx [1]{%
 \ifx #1\expandafter \@firstoftwo
 \else \expandafter \@secondoftwo
 \fi
}%
\providecommand \natexlab [1]{#1}%
\providecommand \enquote  [1]{``#1''}%
\providecommand \bibnamefont  [1]{#1}%
\providecommand \bibfnamefont [1]{#1}%
\providecommand \citenamefont [1]{#1}%
\providecommand \href@noop [0]{\@secondoftwo}%
\providecommand \href [0]{\begingroup \@sanitize@url \@href}%
\providecommand \@href[1]{\@@startlink{#1}\@@href}%
\providecommand \@@href[1]{\endgroup#1\@@endlink}%
\providecommand \@sanitize@url [0]{\catcode `\\12\catcode `\$12\catcode
  `\&12\catcode `\#12\catcode `\^12\catcode `\_12\catcode `\%12\relax}%
\providecommand \@@startlink[1]{}%
\providecommand \@@endlink[0]{}%
\providecommand \url  [0]{\begingroup\@sanitize@url \@url }%
\providecommand \@url [1]{\endgroup\@href {#1}{\urlprefix }}%
\providecommand \urlprefix  [0]{URL }%
\providecommand \Eprint [0]{\href }%
\providecommand \doibase [0]{http://dx.doi.org/}%
\providecommand \selectlanguage [0]{\@gobble}%
\providecommand \bibinfo  [0]{\@secondoftwo}%
\providecommand \bibfield  [0]{\@secondoftwo}%
\providecommand \translation [1]{[#1]}%
\providecommand \BibitemOpen [0]{}%
\providecommand \bibitemStop [0]{}%
\providecommand \bibitemNoStop [0]{.\EOS\space}%
\providecommand \EOS [0]{\spacefactor3000\relax}%
\providecommand \BibitemShut  [1]{\csname bibitem#1\endcsname}%
\let\auto@bib@innerbib\@empty
\bibitem [{\citenamefont {H{\"a}nsch}\ and\ \citenamefont
  {Schawlow}(1975)}]{hansch1975}%
  \BibitemOpen
  \bibfield  {author} {\bibinfo {author} {\bibfnamefont {T.}~\bibnamefont
  {H{\"a}nsch}}\ and\ \bibinfo {author} {\bibfnamefont {A.}~\bibnamefont
  {Schawlow}},\ }\href {\doibase
  http://dx.doi.org/10.1016/0030-4018(75)90159-5} {\bibfield  {journal}
  {\bibinfo  {journal} {Opt. Comm.}\ }\textbf {\bibinfo {volume} {13}},\
  \bibinfo {pages} {68 } (\bibinfo {year} {1975})}\BibitemShut {NoStop}%
\bibitem [{\citenamefont {Wineland}\ and\ \citenamefont
  {Itano}(1979)}]{wineland1979}%
  \BibitemOpen
  \bibfield  {author} {\bibinfo {author} {\bibfnamefont {D.~J.}\ \bibnamefont
  {Wineland}}\ and\ \bibinfo {author} {\bibfnamefont {W.~M.}\ \bibnamefont
  {Itano}},\ }\href {\doibase 10.1103/PhysRevA.20.1521} {\bibfield  {journal}
  {\bibinfo  {journal} {Phys. Rev. A}\ }\textbf {\bibinfo {volume} {20}},\
  \bibinfo {pages} {1521} (\bibinfo {year} {1979})}\BibitemShut {NoStop}%
\bibitem [{\citenamefont {Raab}\ \emph {et~al.}(1987)\citenamefont {Raab},
  \citenamefont {Prentiss}, \citenamefont {Cable}, \citenamefont {Chu},\ and\
  \citenamefont {Pritchard}}]{raab1987}%
  \BibitemOpen
  \bibfield  {author} {\bibinfo {author} {\bibfnamefont {E.~L.}\ \bibnamefont
  {Raab}}, \bibinfo {author} {\bibfnamefont {M.}~\bibnamefont {Prentiss}},
  \bibinfo {author} {\bibfnamefont {A.}~\bibnamefont {Cable}}, \bibinfo
  {author} {\bibfnamefont {S.}~\bibnamefont {Chu}}, \ and\ \bibinfo {author}
  {\bibfnamefont {D.~E.}\ \bibnamefont {Pritchard}},\ }\href {\doibase
  10.1103/PhysRevLett.59.2631} {\bibfield  {journal} {\bibinfo  {journal}
  {Phys. Rev. Lett.}\ }\textbf {\bibinfo {volume} {59}},\ \bibinfo {pages}
  {2631} (\bibinfo {year} {1987})}\BibitemShut {NoStop}%
\bibitem [{\citenamefont {Grimm}\ \emph {et~al.}(2000)\citenamefont {Grimm},
  \citenamefont {Weidem{\"u}ller},\ and\ \citenamefont
  {Ovchinnikov}}]{grimm2000}%
  \BibitemOpen
  \bibfield  {author} {\bibinfo {author} {\bibfnamefont {R.}~\bibnamefont
  {Grimm}}, \bibinfo {author} {\bibfnamefont {M.}~\bibnamefont
  {Weidem{\"u}ller}}, \ and\ \bibinfo {author} {\bibfnamefont {Y.~B.}\
  \bibnamefont {Ovchinnikov}},\ }in\ \href@noop {} {\emph {\bibinfo {booktitle}
  {Adv. At. Mol. Opt. Phys.}}},\ Vol.~\bibinfo {volume} {42},\ \bibinfo
  {editor} {edited by\ \bibinfo {editor} {\bibfnamefont {B.~B.}\ \bibnamefont
  {Walther}}\ and\ \bibinfo {editor} {\bibnamefont {Herbert}}}\ (\bibinfo
  {publisher} {{Academic Press}},\ \bibinfo {year} {2000})\ pp.\ \bibinfo
  {pages} {95--170}\BibitemShut {NoStop}%
\bibitem [{\citenamefont {Fioretti}\ \emph {et~al.}(1998)\citenamefont
  {Fioretti}, \citenamefont {Comparat}, \citenamefont {Crubellier},
  \citenamefont {Dulieu}, \citenamefont {Masnou-Seeuws},\ and\ \citenamefont
  {Pillet}}]{fioretti1998}%
  \BibitemOpen
  \bibfield  {author} {\bibinfo {author} {\bibfnamefont {A.}~\bibnamefont
  {Fioretti}}, \bibinfo {author} {\bibfnamefont {D.}~\bibnamefont {Comparat}},
  \bibinfo {author} {\bibfnamefont {A.}~\bibnamefont {Crubellier}}, \bibinfo
  {author} {\bibfnamefont {O.}~\bibnamefont {Dulieu}}, \bibinfo {author}
  {\bibfnamefont {F.}~\bibnamefont {Masnou-Seeuws}}, \ and\ \bibinfo {author}
  {\bibfnamefont {P.}~\bibnamefont {Pillet}},\ }\href@noop {} {\bibfield
  {journal} {\bibinfo  {journal} {Phys. Rev. Lett.}\ }\textbf {\bibinfo
  {volume} {80}},\ \bibinfo {pages} {4402} (\bibinfo {year}
  {1998})}\BibitemShut {NoStop}%
\bibitem [{\citenamefont {Viteau}\ \emph
  {et~al.}(2008{\natexlab{a}})\citenamefont {Viteau}, \citenamefont {Chotia},
  \citenamefont {Allegrini}, \citenamefont {Bouloufa}, \citenamefont {Dulieu},
  \citenamefont {Comparat},\ and\ \citenamefont {Pillet}}]{viteau2008}%
  \BibitemOpen
  \bibfield  {author} {\bibinfo {author} {\bibfnamefont {M.}~\bibnamefont
  {Viteau}}, \bibinfo {author} {\bibfnamefont {A.}~\bibnamefont {Chotia}},
  \bibinfo {author} {\bibfnamefont {M.}~\bibnamefont {Allegrini}}, \bibinfo
  {author} {\bibfnamefont {N.}~\bibnamefont {Bouloufa}}, \bibinfo {author}
  {\bibfnamefont {O.}~\bibnamefont {Dulieu}}, \bibinfo {author} {\bibfnamefont
  {D.}~\bibnamefont {Comparat}}, \ and\ \bibinfo {author} {\bibfnamefont
  {P.}~\bibnamefont {Pillet}},\ }\href@noop {} {\bibfield  {journal} {\bibinfo
  {journal} {Science}\ }\textbf {\bibinfo {volume} {321}},\ \bibinfo {pages}
  {232} (\bibinfo {year} {2008}{\natexlab{a}})}\BibitemShut {NoStop}%
\bibitem [{\citenamefont {Deiglmayr}\ \emph {et~al.}(2008)\citenamefont
  {Deiglmayr}, \citenamefont {Grochola}, \citenamefont {Repp}, \citenamefont
  {M{\"o}rtlbauer}, \citenamefont {Gl{\"u}ck}, \citenamefont {Lange},
  \citenamefont {Dulieu}, \citenamefont {Wester},\ and\ \citenamefont
  {Weidem{\"u}ller}}]{deiglmayr2008}%
  \BibitemOpen
  \bibfield  {author} {\bibinfo {author} {\bibfnamefont {J.}~\bibnamefont
  {Deiglmayr}}, \bibinfo {author} {\bibfnamefont {A.}~\bibnamefont {Grochola}},
  \bibinfo {author} {\bibfnamefont {M.}~\bibnamefont {Repp}}, \bibinfo {author}
  {\bibfnamefont {K.}~\bibnamefont {M{\"o}rtlbauer}}, \bibinfo {author}
  {\bibfnamefont {C.}~\bibnamefont {Gl{\"u}ck}}, \bibinfo {author}
  {\bibfnamefont {J.}~\bibnamefont {Lange}}, \bibinfo {author} {\bibfnamefont
  {O.}~\bibnamefont {Dulieu}}, \bibinfo {author} {\bibfnamefont
  {R.}~\bibnamefont {Wester}}, \ and\ \bibinfo {author} {\bibfnamefont
  {M.}~\bibnamefont {Weidem{\"u}ller}},\ }\href@noop {} {\bibfield  {journal}
  {\bibinfo  {journal} {Phys. Rev. Lett.}\ }\textbf {\bibinfo {volume} {101}},\
  \bibinfo {pages} {133004} (\bibinfo {year} {2008})}\BibitemShut {NoStop}%
\bibitem [{\citenamefont {Ni}\ \emph {et~al.}(2008)\citenamefont {Ni},
  \citenamefont {Ospelkaus}, \citenamefont {{de Miranda}}, \citenamefont
  {Pe'er}, \citenamefont {Neyenhuis}, \citenamefont {Zirbel}, \citenamefont
  {Kotochigova}, \citenamefont {Julienne}, \citenamefont {Jin},\ and\
  \citenamefont {Ye}}]{ni2008}%
  \BibitemOpen
  \bibfield  {author} {\bibinfo {author} {\bibfnamefont {K.-K.}\ \bibnamefont
  {Ni}}, \bibinfo {author} {\bibfnamefont {S.}~\bibnamefont {Ospelkaus}},
  \bibinfo {author} {\bibfnamefont {M.~H.~G.}\ \bibnamefont {{de Miranda}}},
  \bibinfo {author} {\bibfnamefont {A.}~\bibnamefont {Pe'er}}, \bibinfo
  {author} {\bibfnamefont {B.}~\bibnamefont {Neyenhuis}}, \bibinfo {author}
  {\bibfnamefont {J.~J.}\ \bibnamefont {Zirbel}}, \bibinfo {author}
  {\bibfnamefont {S.}~\bibnamefont {Kotochigova}}, \bibinfo {author}
  {\bibfnamefont {P.~S.}\ \bibnamefont {Julienne}}, \bibinfo {author}
  {\bibfnamefont {D.~S.}\ \bibnamefont {Jin}}, \ and\ \bibinfo {author}
  {\bibfnamefont {J.}~\bibnamefont {Ye}},\ }\href@noop {} {\bibfield  {journal}
  {\bibinfo  {journal} {Science}\ }\textbf {\bibinfo {volume} {322}},\ \bibinfo
  {pages} {231} (\bibinfo {year} {2008})}\BibitemShut {NoStop}%
\bibitem [{\citenamefont {Danzl}\ \emph {et~al.}(2010)\citenamefont {Danzl},
  \citenamefont {Mark}, \citenamefont {Haller}, \citenamefont {Gustavsson},
  \citenamefont {Hart}, \citenamefont {Aldegunde}, \citenamefont {Hutson},\
  and\ \citenamefont {N{\"a}gerl}}]{danzl2010}%
  \BibitemOpen
  \bibfield  {author} {\bibinfo {author} {\bibfnamefont {J.~G.}\ \bibnamefont
  {Danzl}}, \bibinfo {author} {\bibfnamefont {M.~J.}\ \bibnamefont {Mark}},
  \bibinfo {author} {\bibfnamefont {E.}~\bibnamefont {Haller}}, \bibinfo
  {author} {\bibfnamefont {M.}~\bibnamefont {Gustavsson}}, \bibinfo {author}
  {\bibfnamefont {R.}~\bibnamefont {Hart}}, \bibinfo {author} {\bibfnamefont
  {J.}~\bibnamefont {Aldegunde}}, \bibinfo {author} {\bibfnamefont {J.~M.}\
  \bibnamefont {Hutson}}, \ and\ \bibinfo {author} {\bibfnamefont {H.-C.}\
  \bibnamefont {N{\"a}gerl}},\ }\href@noop {} {\bibfield  {journal} {\bibinfo
  {journal} {Nat. Phys.}\ }\textbf {\bibinfo {volume} {6}},\ \bibinfo {pages}
  {265} (\bibinfo {year} {2010})}\BibitemShut {NoStop}%
\bibitem [{\citenamefont {Ospelkaus}\ \emph {et~al.}(2010)\citenamefont
  {Ospelkaus}, \citenamefont {Ni}, \citenamefont {Wang}, \citenamefont {{de
  Miranda}}, \citenamefont {Neyenhuis}, \citenamefont {Qu{\'e}m{\'e}ner},
  \citenamefont {Julienne}, \citenamefont {Bohn}, \citenamefont {Jin},\ and\
  \citenamefont {Ye}}]{ospelkaus2010}%
  \BibitemOpen
  \bibfield  {author} {\bibinfo {author} {\bibfnamefont {S.}~\bibnamefont
  {Ospelkaus}}, \bibinfo {author} {\bibfnamefont {K.-K.}\ \bibnamefont {Ni}},
  \bibinfo {author} {\bibfnamefont {D.}~\bibnamefont {Wang}}, \bibinfo {author}
  {\bibfnamefont {M.~H.~G.}\ \bibnamefont {{de Miranda}}}, \bibinfo {author}
  {\bibfnamefont {B.}~\bibnamefont {Neyenhuis}}, \bibinfo {author}
  {\bibfnamefont {G.}~\bibnamefont {Qu{\'e}m{\'e}ner}}, \bibinfo {author}
  {\bibfnamefont {P.~S.}\ \bibnamefont {Julienne}}, \bibinfo {author}
  {\bibfnamefont {J.~L.}\ \bibnamefont {Bohn}}, \bibinfo {author}
  {\bibfnamefont {D.~S.}\ \bibnamefont {Jin}}, \ and\ \bibinfo {author}
  {\bibfnamefont {J.}~\bibnamefont {Ye}},\ }\href@noop {} {\bibfield  {journal}
  {\bibinfo  {journal} {Science}\ }\textbf {\bibinfo {volume} {327}},\ \bibinfo
  {pages} {853} (\bibinfo {year} {2010})}\BibitemShut {NoStop}%
\bibitem [{\citenamefont {Knoop}\ \emph {et~al.}(2010)\citenamefont {Knoop},
  \citenamefont {Ferlaino}, \citenamefont {Berninger}, \citenamefont {Mark},
  \citenamefont {N{\"a}gerl}, \citenamefont {Grimm}, \citenamefont {D'Incao},\
  and\ \citenamefont {Esry}}]{knoop2010}%
  \BibitemOpen
  \bibfield  {author} {\bibinfo {author} {\bibfnamefont {S.}~\bibnamefont
  {Knoop}}, \bibinfo {author} {\bibfnamefont {F.}~\bibnamefont {Ferlaino}},
  \bibinfo {author} {\bibfnamefont {M.}~\bibnamefont {Berninger}}, \bibinfo
  {author} {\bibfnamefont {M.}~\bibnamefont {Mark}}, \bibinfo {author}
  {\bibfnamefont {H.-C.}\ \bibnamefont {N{\"a}gerl}}, \bibinfo {author}
  {\bibfnamefont {R.}~\bibnamefont {Grimm}}, \bibinfo {author} {\bibfnamefont
  {J.~P.}\ \bibnamefont {D'Incao}}, \ and\ \bibinfo {author} {\bibfnamefont
  {B.~D.}\ \bibnamefont {Esry}},\ }\href@noop {} {\bibfield  {journal}
  {\bibinfo  {journal} {Phys. Rev. Lett.}\ }\textbf {\bibinfo {volume} {104}},\
  \bibinfo {pages} {053201} (\bibinfo {year} {2010})}\BibitemShut {NoStop}%
\bibitem [{\citenamefont {McDonald}\ \emph {et~al.}(2016)\citenamefont
  {McDonald}, \citenamefont {McGuyer}, \citenamefont {Apfelbeck}, \citenamefont
  {Lee}, \citenamefont {Majewska}, \citenamefont {Moszynski},\ and\
  \citenamefont {Zelevinsky}}]{mcDonald2016}%
  \BibitemOpen
  \bibfield  {author} {\bibinfo {author} {\bibfnamefont {M.}~\bibnamefont
  {McDonald}}, \bibinfo {author} {\bibfnamefont {B.~H.}\ \bibnamefont
  {McGuyer}}, \bibinfo {author} {\bibfnamefont {F.}~\bibnamefont {Apfelbeck}},
  \bibinfo {author} {\bibfnamefont {C.-H.}\ \bibnamefont {Lee}}, \bibinfo
  {author} {\bibfnamefont {I.}~\bibnamefont {Majewska}}, \bibinfo {author}
  {\bibfnamefont {R.}~\bibnamefont {Moszynski}}, \ and\ \bibinfo {author}
  {\bibfnamefont {T.}~\bibnamefont {Zelevinsky}},\ }\href@noop {} {\bibfield
  {journal} {\bibinfo  {journal} {Nature}\ }\textbf {\bibinfo {volume} {534}},\
  \bibinfo {pages} {122} (\bibinfo {year} {2016})}\BibitemShut {NoStop}%
\bibitem [{\citenamefont {Gallagher}\ and\ \citenamefont
  {Pillet}(2008)}]{gallagher2008}%
  \BibitemOpen
  \bibfield  {author} {\bibinfo {author} {\bibfnamefont {T.~F.}\ \bibnamefont
  {Gallagher}}\ and\ \bibinfo {author} {\bibfnamefont {P.}~\bibnamefont
  {Pillet}},\ }in\ \href
  {http://www.sciencedirect.com/science/article/pii/S1049250X0800013X} {\emph
  {\bibinfo {booktitle} {Adv. At. Mol. Opt. Phys.}}},\ Vol.~\bibinfo {volume}
  {56},\ \bibinfo {editor} {edited by\ \bibinfo {editor} {\bibfnamefont
  {E.}~\bibnamefont {Arimondo}}}\ (\bibinfo  {publisher} {Academic Press},\
  \bibinfo {year} {2008})\ pp.\ \bibinfo {pages} {161--218}\BibitemShut
  {NoStop}%
\bibitem [{\citenamefont {Boisseau}\ \emph {et~al.}(2002)\citenamefont
  {Boisseau}, \citenamefont {Simbotin},\ and\ \citenamefont
  {C\^ot\'e}}]{boisseau2002}%
  \BibitemOpen
  \bibfield  {author} {\bibinfo {author} {\bibfnamefont {C.}~\bibnamefont
  {Boisseau}}, \bibinfo {author} {\bibfnamefont {I.}~\bibnamefont {Simbotin}},
  \ and\ \bibinfo {author} {\bibfnamefont {R.}~\bibnamefont {C\^ot\'e}},\
  }\href {\doibase 10.1103/PhysRevLett.88.133004} {\bibfield  {journal}
  {\bibinfo  {journal} {Phys. Rev. Lett.}\ }\textbf {\bibinfo {volume} {88}},\
  \bibinfo {pages} {133004} (\bibinfo {year} {2002})}\BibitemShut {NoStop}%
\bibitem [{\citenamefont {Schwettmann}\ \emph {et~al.}(2007)\citenamefont
  {Schwettmann}, \citenamefont {Overstreet}, \citenamefont {Tallant},\ and\
  \citenamefont {Shaffer}}]{schwettmann2007}%
  \BibitemOpen
  \bibfield  {author} {\bibinfo {author} {\bibfnamefont {A.}~\bibnamefont
  {Schwettmann}}, \bibinfo {author} {\bibfnamefont {K.~R.}\ \bibnamefont
  {Overstreet}}, \bibinfo {author} {\bibfnamefont {J.}~\bibnamefont {Tallant}},
  \ and\ \bibinfo {author} {\bibfnamefont {J.~P.}\ \bibnamefont {Shaffer}},\
  }\href@noop {} {\bibfield  {journal} {\bibinfo  {journal} {J. Mod. Opt.}\
  }\textbf {\bibinfo {volume} {54}},\ \bibinfo {pages} {2551} (\bibinfo {year}
  {2007})}\BibitemShut {NoStop}%
\bibitem [{\citenamefont {Kiffner}\ \emph {et~al.}(2012)\citenamefont
  {Kiffner}, \citenamefont {Park}, \citenamefont {Li},\ and\ \citenamefont
  {Gallagher}}]{kiffner2012}%
  \BibitemOpen
  \bibfield  {author} {\bibinfo {author} {\bibfnamefont {M.}~\bibnamefont
  {Kiffner}}, \bibinfo {author} {\bibfnamefont {H.}~\bibnamefont {Park}},
  \bibinfo {author} {\bibfnamefont {W.}~\bibnamefont {Li}}, \ and\ \bibinfo
  {author} {\bibfnamefont {T.~F.}\ \bibnamefont {Gallagher}},\ }\href {\doibase
  10.1103/PhysRevA.86.031401} {\bibfield  {journal} {\bibinfo  {journal} {Phys.
  Rev. A}\ }\textbf {\bibinfo {volume} {86}},\ \bibinfo {pages} {031401}
  (\bibinfo {year} {2012})}\BibitemShut {NoStop}%
\bibitem [{\citenamefont {Marcassa}\ and\ \citenamefont
  {Shaffer}(2014)}]{marcassa2014}%
  \BibitemOpen
  \bibfield  {author} {\bibinfo {author} {\bibfnamefont {L.~G.}\ \bibnamefont
  {Marcassa}}\ and\ \bibinfo {author} {\bibfnamefont {J.~P.}\ \bibnamefont
  {Shaffer}},\ }in\ \href@noop {} {\emph {\bibinfo {booktitle} {Adv. At. Mol.
  Opt. Phys.}}},\ Vol.~\bibinfo {volume} {63},\ \bibinfo {editor} {edited by\
  \bibinfo {editor} {\bibfnamefont {P.~R.~B.}\ \bibnamefont {{Ennio
  Arimondo}}}\ and\ \bibinfo {editor} {\bibfnamefont {C.~C.}\ \bibnamefont
  {Lin}}}\ (\bibinfo  {publisher} {{Academic Press}},\ \bibinfo {year} {2014})\
  pp.\ \bibinfo {pages} {47--133}\BibitemShut {NoStop}%
\bibitem [{\citenamefont {Lefebvre-Brion}\ and\ \citenamefont
  {Field}(2004)}]{lefebvre2004}%
  \BibitemOpen
  \bibfield  {author} {\bibinfo {author} {\bibfnamefont {H.}~\bibnamefont
  {Lefebvre-Brion}}\ and\ \bibinfo {author} {\bibfnamefont {R.~W.}\
  \bibnamefont {Field}},\ }\href@noop {} {\emph {\bibinfo {title} {The
  {{Spectra}} and {{Dynamics}} of {{Diatomic Molecules}}}}}\ (\bibinfo
  {publisher} {{Academic Press}},\ \bibinfo {address} {San Diego},\ \bibinfo
  {year} {2004})\BibitemShut {NoStop}%
\bibitem [{\citenamefont {Farooqi}\ \emph {et~al.}(2003)\citenamefont
  {Farooqi}, \citenamefont {Tong}, \citenamefont {Krishnan}, \citenamefont
  {Stanojevic}, \citenamefont {Zhang}, \citenamefont {Ensher}, \citenamefont
  {Estrin}, \citenamefont {Boisseau}, \citenamefont {C\^ot\'e}, \citenamefont
  {Eyler},\ and\ \citenamefont {Gould}}]{farooqi2003}%
  \BibitemOpen
  \bibfield  {author} {\bibinfo {author} {\bibfnamefont {S.~M.}\ \bibnamefont
  {Farooqi}}, \bibinfo {author} {\bibfnamefont {D.}~\bibnamefont {Tong}},
  \bibinfo {author} {\bibfnamefont {S.}~\bibnamefont {Krishnan}}, \bibinfo
  {author} {\bibfnamefont {J.}~\bibnamefont {Stanojevic}}, \bibinfo {author}
  {\bibfnamefont {Y.~P.}\ \bibnamefont {Zhang}}, \bibinfo {author}
  {\bibfnamefont {J.~R.}\ \bibnamefont {Ensher}}, \bibinfo {author}
  {\bibfnamefont {A.~S.}\ \bibnamefont {Estrin}}, \bibinfo {author}
  {\bibfnamefont {C.}~\bibnamefont {Boisseau}}, \bibinfo {author}
  {\bibfnamefont {R.}~\bibnamefont {C\^ot\'e}}, \bibinfo {author}
  {\bibfnamefont {E.~E.}\ \bibnamefont {Eyler}}, \ and\ \bibinfo {author}
  {\bibfnamefont {P.~L.}\ \bibnamefont {Gould}},\ }\href {\doibase
  10.1103/PhysRevLett.91.183002} {\bibfield  {journal} {\bibinfo  {journal}
  {Phys. Rev. Lett.}\ }\textbf {\bibinfo {volume} {91}},\ \bibinfo {pages}
  {183002} (\bibinfo {year} {2003})}\BibitemShut {NoStop}%
\bibitem [{\citenamefont {Overstreet}\ \emph {et~al.}(2009)\citenamefont
  {Overstreet}, \citenamefont {Schwettmann}, \citenamefont {Tallant},
  \citenamefont {Booth},\ and\ \citenamefont {Shaffer}}]{overstreet2009}%
  \BibitemOpen
  \bibfield  {author} {\bibinfo {author} {\bibfnamefont {K.~R.}\ \bibnamefont
  {Overstreet}}, \bibinfo {author} {\bibfnamefont {A.}~\bibnamefont
  {Schwettmann}}, \bibinfo {author} {\bibfnamefont {J.}~\bibnamefont
  {Tallant}}, \bibinfo {author} {\bibfnamefont {D.}~\bibnamefont {Booth}}, \
  and\ \bibinfo {author} {\bibfnamefont {J.~P.}\ \bibnamefont {Shaffer}},\
  }\href {\doibase 10.1038/nphys1307} {\bibfield  {journal} {\bibinfo
  {journal} {Nat. Phys.}\ }\textbf {\bibinfo {volume} {5}},\ \bibinfo {pages}
  {581} (\bibinfo {year} {2009})}\BibitemShut {NoStop}%
\bibitem [{\citenamefont {Deiglmayr}\ \emph {et~al.}(2014)\citenamefont
  {Deiglmayr}, \citenamefont {Sa{\ss}mannshausen}, \citenamefont {Pillet},\
  and\ \citenamefont {Merkt}}]{deiglmayr2014}%
  \BibitemOpen
  \bibfield  {author} {\bibinfo {author} {\bibfnamefont {J.}~\bibnamefont
  {Deiglmayr}}, \bibinfo {author} {\bibfnamefont {H.}~\bibnamefont
  {Sa{\ss}mannshausen}}, \bibinfo {author} {\bibfnamefont {P.}~\bibnamefont
  {Pillet}}, \ and\ \bibinfo {author} {\bibfnamefont {F.}~\bibnamefont
  {Merkt}},\ }\href@noop {} {\bibfield  {journal} {\bibinfo  {journal} {Phys.
  Rev. Lett.}\ }\textbf {\bibinfo {volume} {113}},\ \bibinfo {pages} {193001}
  (\bibinfo {year} {2014})}\BibitemShut {NoStop}%
\bibitem [{\citenamefont {Sa\ss{}mannshausen}\ \emph
  {et~al.}(2015)\citenamefont {Sa\ss{}mannshausen}, \citenamefont {Merkt},\
  and\ \citenamefont {Deiglmayr}}]{sassmannshausen2015}%
  \BibitemOpen
  \bibfield  {author} {\bibinfo {author} {\bibfnamefont {H.}~\bibnamefont
  {Sa\ss{}mannshausen}}, \bibinfo {author} {\bibfnamefont {F.}~\bibnamefont
  {Merkt}}, \ and\ \bibinfo {author} {\bibfnamefont {J.}~\bibnamefont
  {Deiglmayr}},\ }\href {\doibase 10.1103/PhysRevA.92.032505} {\bibfield
  {journal} {\bibinfo  {journal} {Phys. Rev. A}\ }\textbf {\bibinfo {volume}
  {92}},\ \bibinfo {pages} {032505} (\bibinfo {year} {2015})}\BibitemShut
  {NoStop}%
\bibitem [{\citenamefont {Goy}\ \emph {et~al.}(1982)\citenamefont {Goy},
  \citenamefont {Raimond}, \citenamefont {Vitrant},\ and\ \citenamefont
  {Haroche}}]{goy1982}%
  \BibitemOpen
  \bibfield  {author} {\bibinfo {author} {\bibfnamefont {P.}~\bibnamefont
  {Goy}}, \bibinfo {author} {\bibfnamefont {J.~M.}\ \bibnamefont {Raimond}},
  \bibinfo {author} {\bibfnamefont {G.}~\bibnamefont {Vitrant}}, \ and\
  \bibinfo {author} {\bibfnamefont {S.}~\bibnamefont {Haroche}},\ }\href
  {\doibase 10.1103/PhysRevA.26.2733} {\bibfield  {journal} {\bibinfo
  {journal} {Phys. Rev. A}\ }\textbf {\bibinfo {volume} {26}},\ \bibinfo
  {pages} {2733} (\bibinfo {year} {1982})}\BibitemShut {NoStop}%
\bibitem [{\citenamefont {Le~Roy}(1973)}]{leroy1973}%
  \BibitemOpen
  \bibfield  {author} {\bibinfo {author} {\bibfnamefont {R.~J.}\ \bibnamefont
  {Le~Roy}},\ }in\ \href
  {http://pubs.rsc.org/en/content/chapter/bk9780851865065-00113/978-0-85186-506-5#!divabstract}
  {\emph {\bibinfo {booktitle} {Molecular Spectroscopy}}},\ Vol.~\bibinfo
  {volume} {1},\ \bibinfo {editor} {edited by\ \bibinfo {editor} {\bibfnamefont
  {R.~F.}\ \bibnamefont {Barrow}}, \bibinfo {editor} {\bibfnamefont {D.~A.}\
  \bibnamefont {Long}}, \ and\ \bibinfo {editor} {\bibfnamefont {D.~J.}\
  \bibnamefont {Millen}}}\ (\bibinfo  {publisher} {Royal Society of
  Chemistry},\ \bibinfo {address} {Cambridge},\ \bibinfo {year} {1973})\ pp.\
  \bibinfo {pages} {113--176}\BibitemShut {NoStop}%
\bibitem [{\citenamefont {Kiffner}\ \emph {et~al.}(2014)\citenamefont
  {Kiffner}, \citenamefont {Huo}, \citenamefont {Li},\ and\ \citenamefont
  {Jaksch}}]{kiffner2014}%
  \BibitemOpen
  \bibfield  {author} {\bibinfo {author} {\bibfnamefont {M.}~\bibnamefont
  {Kiffner}}, \bibinfo {author} {\bibfnamefont {M.}~\bibnamefont {Huo}},
  \bibinfo {author} {\bibfnamefont {W.}~\bibnamefont {Li}}, \ and\ \bibinfo
  {author} {\bibfnamefont {D.}~\bibnamefont {Jaksch}},\ }\href@noop {}
  {\bibfield  {journal} {\bibinfo  {journal} {Phys. Rev. A}\ }\textbf {\bibinfo
  {volume} {89}},\ \bibinfo {pages} {052717} (\bibinfo {year}
  {2014})}\BibitemShut {NoStop}%
\bibitem [{\citenamefont {Samboy}\ \emph {et~al.}(2011)\citenamefont {Samboy},
  \citenamefont {Stanojevic},\ and\ \citenamefont {C{\^o}t{\'e}}}]{samboy2011}%
  \BibitemOpen
  \bibfield  {author} {\bibinfo {author} {\bibfnamefont {N.}~\bibnamefont
  {Samboy}}, \bibinfo {author} {\bibfnamefont {J.}~\bibnamefont {Stanojevic}},
  \ and\ \bibinfo {author} {\bibfnamefont {R.}~\bibnamefont {C{\^o}t{\'e}}},\
  }\href@noop {} {\bibfield  {journal} {\bibinfo  {journal} {Phys. Rev. A}\
  }\textbf {\bibinfo {volume} {83}},\ \bibinfo {pages} {050501} (\bibinfo
  {year} {2011})}\BibitemShut {NoStop}%
\bibitem [{\citenamefont {Sa\ss{}mannshausen}\ \emph
  {et~al.}(2016)\citenamefont {Sa\ss{}mannshausen}, \citenamefont {Deiglmayr},\
  and\ \citenamefont {Merkt}}]{sassmannshausen2016}%
  \BibitemOpen
  \bibfield  {author} {\bibinfo {author} {\bibfnamefont {H.}~\bibnamefont
  {Sa\ss{}mannshausen}}, \bibinfo {author} {\bibfnamefont {J.}~\bibnamefont
  {Deiglmayr}}, \ and\ \bibinfo {author} {\bibfnamefont {F.}~\bibnamefont
  {Merkt}},\ }\href@noop {} {\bibfield  {journal} {\bibinfo  {journal}
  {submitted to Europ. Phys. J. Spec. Top.}\ } (\bibinfo {year}
  {2016})}\BibitemShut {NoStop}%
\bibitem [{\citenamefont {Lukin}\ \emph {et~al.}(2001)\citenamefont {Lukin},
  \citenamefont {Fleischhauer}, \citenamefont {Cote}, \citenamefont {Duan},
  \citenamefont {Jaksch}, \citenamefont {Cirac},\ and\ \citenamefont
  {Zoller}}]{lukin2001}%
  \BibitemOpen
  \bibfield  {author} {\bibinfo {author} {\bibfnamefont {M.~D.}\ \bibnamefont
  {Lukin}}, \bibinfo {author} {\bibfnamefont {M.}~\bibnamefont {Fleischhauer}},
  \bibinfo {author} {\bibfnamefont {R.}~\bibnamefont {Cote}}, \bibinfo {author}
  {\bibfnamefont {L.~M.}\ \bibnamefont {Duan}}, \bibinfo {author}
  {\bibfnamefont {D.}~\bibnamefont {Jaksch}}, \bibinfo {author} {\bibfnamefont
  {J.~I.}\ \bibnamefont {Cirac}}, \ and\ \bibinfo {author} {\bibfnamefont
  {P.}~\bibnamefont {Zoller}},\ }\href@noop {} {\bibfield  {journal} {\bibinfo
  {journal} {Phys. Rev. Lett.}\ }\textbf {\bibinfo {volume} {87}},\ \bibinfo
  {pages} {037901} (\bibinfo {year} {2001})}\BibitemShut {NoStop}%
\bibitem [{\citenamefont {Reinhard}\ \emph {et~al.}(2008)\citenamefont
  {Reinhard}, \citenamefont {Younge}, \citenamefont {Liebisch}, \citenamefont
  {Knuffman}, \citenamefont {Berman},\ and\ \citenamefont
  {Raithel}}]{reinhard2008}%
  \BibitemOpen
  \bibfield  {author} {\bibinfo {author} {\bibfnamefont {A.}~\bibnamefont
  {Reinhard}}, \bibinfo {author} {\bibfnamefont {K.~C.}\ \bibnamefont
  {Younge}}, \bibinfo {author} {\bibfnamefont {T.~C.}\ \bibnamefont
  {Liebisch}}, \bibinfo {author} {\bibfnamefont {B.}~\bibnamefont {Knuffman}},
  \bibinfo {author} {\bibfnamefont {P.~R.}\ \bibnamefont {Berman}}, \ and\
  \bibinfo {author} {\bibfnamefont {G.}~\bibnamefont {Raithel}},\ }\href
  {\doibase 10.1103/PhysRevLett.100.233201} {\bibfield  {journal} {\bibinfo
  {journal} {Phys. Rev. Lett.}\ }\textbf {\bibinfo {volume} {100}},\ \bibinfo
  {pages} {233201} (\bibinfo {year} {2008})}\BibitemShut {NoStop}%
\bibitem [{\citenamefont {Thorsheim}\ \emph {et~al.}(1987)\citenamefont
  {Thorsheim}, \citenamefont {Weiner},\ and\ \citenamefont
  {Julienne}}]{thorsheim1987}%
  \BibitemOpen
  \bibfield  {author} {\bibinfo {author} {\bibfnamefont {H.~R.}\ \bibnamefont
  {Thorsheim}}, \bibinfo {author} {\bibfnamefont {J.}~\bibnamefont {Weiner}}, \
  and\ \bibinfo {author} {\bibfnamefont {P.~S.}\ \bibnamefont {Julienne}},\
  }\href@noop {} {\bibfield  {journal} {\bibinfo  {journal} {Phys. Rev. Lett.}\
  }\textbf {\bibinfo {volume} {58}},\ \bibinfo {pages} {2420} (\bibinfo {year}
  {1987})}\BibitemShut {NoStop}%
\bibitem [{\citenamefont {Bambini}\ \emph {et~al.}(1994)\citenamefont
  {Bambini}, \citenamefont {Berman}, \citenamefont {Buffa}, \citenamefont
  {Robinson},\ and\ \citenamefont {Matera}}]{bambini1994}%
  \BibitemOpen
  \bibfield  {author} {\bibinfo {author} {\bibfnamefont {A.}~\bibnamefont
  {Bambini}}, \bibinfo {author} {\bibfnamefont {P.~R.}\ \bibnamefont {Berman}},
  \bibinfo {author} {\bibfnamefont {R.}~\bibnamefont {Buffa}}, \bibinfo
  {author} {\bibfnamefont {E.~J.}\ \bibnamefont {Robinson}}, \ and\ \bibinfo
  {author} {\bibfnamefont {M.}~\bibnamefont {Matera}},\ }\href {\doibase
  http://dx.doi.org/10.1016/0370-1573(94)90025-6} {\bibfield  {journal}
  {\bibinfo  {journal} {Physics Reports}\ }\textbf {\bibinfo {volume} {238}},\
  \bibinfo {pages} {245 } (\bibinfo {year} {1994})}\BibitemShut {NoStop}%
\bibitem [{\citenamefont {Sa{\ss}mannshausen}\ \emph
  {et~al.}(2013)\citenamefont {Sa{\ss}mannshausen}, \citenamefont {Merkt},\
  and\ \citenamefont {Deiglmayr}}]{sasmannshausen2013}%
  \BibitemOpen
  \bibfield  {author} {\bibinfo {author} {\bibfnamefont {H.}~\bibnamefont
  {Sa{\ss}mannshausen}}, \bibinfo {author} {\bibfnamefont {F.}~\bibnamefont
  {Merkt}}, \ and\ \bibinfo {author} {\bibfnamefont {J.}~\bibnamefont
  {Deiglmayr}},\ }\href {\doibase 10.1103/PhysRevA.87.032519} {\bibfield
  {journal} {\bibinfo  {journal} {Phys. Rev. A}\ }\textbf {\bibinfo {volume}
  {87}},\ \bibinfo {pages} {032519} (\bibinfo {year} {2013})}\BibitemShut
  {NoStop}%
\bibitem [{\citenamefont {Deiglmayr}\ \emph {et~al.}(2016)\citenamefont
  {Deiglmayr}, \citenamefont {Herburger}, \citenamefont {Sa\ss{}mannshausen},
  \citenamefont {Jansen}, \citenamefont {Schmutz},\ and\ \citenamefont
  {Merkt}}]{deiglmayr2016}%
  \BibitemOpen
  \bibfield  {author} {\bibinfo {author} {\bibfnamefont {J.}~\bibnamefont
  {Deiglmayr}}, \bibinfo {author} {\bibfnamefont {H.}~\bibnamefont
  {Herburger}}, \bibinfo {author} {\bibfnamefont {H.}~\bibnamefont
  {Sa\ss{}mannshausen}}, \bibinfo {author} {\bibfnamefont {P.}~\bibnamefont
  {Jansen}}, \bibinfo {author} {\bibfnamefont {H.}~\bibnamefont {Schmutz}}, \
  and\ \bibinfo {author} {\bibfnamefont {F.}~\bibnamefont {Merkt}},\
  }\href@noop {} {\bibfield  {journal} {\bibinfo  {journal} {Phys. Rev. A}\
  }\textbf {\bibinfo {volume} {93}},\ \bibinfo {pages} {013424} (\bibinfo
  {year} {2016})}\BibitemShut {NoStop}%
\bibitem [{Note1()}]{Note1}%
  \BibitemOpen
  \bibinfo {note} {This assignment is based on the observation that molecular
  Cs$^+_2$ ions are detected on resonance, which we empirically find
  characteristic of Cs$_2$ long-range Rydberg molecules bound by electron-atom
  scattering~\cite {Greene2000,Bendkowsky2009}.}\BibitemShut {Stop}%
\bibitem [{\citenamefont {G\"arttner}\ \emph {et~al.}(2013)\citenamefont
  {G\"arttner}, \citenamefont {Heeg}, \citenamefont {Gasenzer},\ and\
  \citenamefont {Evers}}]{garttner2013}%
  \BibitemOpen
  \bibfield  {author} {\bibinfo {author} {\bibfnamefont {M.}~\bibnamefont
  {G\"arttner}}, \bibinfo {author} {\bibfnamefont {K.~P.}\ \bibnamefont
  {Heeg}}, \bibinfo {author} {\bibfnamefont {T.}~\bibnamefont {Gasenzer}}, \
  and\ \bibinfo {author} {\bibfnamefont {J.}~\bibnamefont {Evers}},\ }\href
  {\doibase 10.1103/PhysRevA.88.043410} {\bibfield  {journal} {\bibinfo
  {journal} {Phys. Rev. A}\ }\textbf {\bibinfo {volume} {88}},\ \bibinfo
  {pages} {043410} (\bibinfo {year} {2013})}\BibitemShut {NoStop}%
\bibitem [{\citenamefont {Lesanovsky}\ and\ \citenamefont
  {Garrahan}(2014)}]{lesanovsky2014}%
  \BibitemOpen
  \bibfield  {author} {\bibinfo {author} {\bibfnamefont {I.}~\bibnamefont
  {Lesanovsky}}\ and\ \bibinfo {author} {\bibfnamefont {J.~P.}\ \bibnamefont
  {Garrahan}},\ }\href {\doibase 10.1103/PhysRevA.90.011603} {\bibfield
  {journal} {\bibinfo  {journal} {Phys. Rev. A}\ }\textbf {\bibinfo {volume}
  {90}},\ \bibinfo {pages} {011603} (\bibinfo {year} {2014})}\BibitemShut
  {NoStop}%
\bibitem [{\citenamefont {Schempp}\ \emph {et~al.}(2014)\citenamefont
  {Schempp}, \citenamefont {G\"unter}, \citenamefont
  {{Robert-de-Saint-Vincent}}, \citenamefont {Hofmann}, \citenamefont {Breyel},
  \citenamefont {Komnik}, \citenamefont {Sch\"onleber}, \citenamefont
  {G\"arttner}, \citenamefont {Evers}, \citenamefont {Whitlock},\ and\
  \citenamefont {Weidem\"uller}}]{schempp2014}%
  \BibitemOpen
  \bibfield  {author} {\bibinfo {author} {\bibfnamefont {H.}~\bibnamefont
  {Schempp}}, \bibinfo {author} {\bibfnamefont {G.}~\bibnamefont {G\"unter}},
  \bibinfo {author} {\bibfnamefont {M.}~\bibnamefont
  {{Robert-de-Saint-Vincent}}}, \bibinfo {author} {\bibfnamefont {C.~S.}\
  \bibnamefont {Hofmann}}, \bibinfo {author} {\bibfnamefont {D.}~\bibnamefont
  {Breyel}}, \bibinfo {author} {\bibfnamefont {A.}~\bibnamefont {Komnik}},
  \bibinfo {author} {\bibfnamefont {D.~W.}\ \bibnamefont {Sch\"onleber}},
  \bibinfo {author} {\bibfnamefont {M.}~\bibnamefont {G\"arttner}}, \bibinfo
  {author} {\bibfnamefont {J.}~\bibnamefont {Evers}}, \bibinfo {author}
  {\bibfnamefont {S.}~\bibnamefont {Whitlock}}, \ and\ \bibinfo {author}
  {\bibfnamefont {M.}~\bibnamefont {Weidem\"uller}},\ }\href {\doibase
  10.1103/PhysRevLett.112.013002} {\bibfield  {journal} {\bibinfo  {journal}
  {Phys. Rev. Lett.}\ }\textbf {\bibinfo {volume} {112}},\ \bibinfo {pages}
  {013002} (\bibinfo {year} {2014})}\BibitemShut {NoStop}%
\bibitem [{\citenamefont {Malossi}\ \emph {et~al.}(2014)\citenamefont
  {Malossi}, \citenamefont {Valado}, \citenamefont {Scotto}, \citenamefont
  {Huillery}, \citenamefont {Pillet}, \citenamefont {Ciampini}, \citenamefont
  {Arimondo},\ and\ \citenamefont {Morsch}}]{malossi2014}%
  \BibitemOpen
  \bibfield  {author} {\bibinfo {author} {\bibfnamefont {N.}~\bibnamefont
  {Malossi}}, \bibinfo {author} {\bibfnamefont {M.~M.}\ \bibnamefont {Valado}},
  \bibinfo {author} {\bibfnamefont {S.}~\bibnamefont {Scotto}}, \bibinfo
  {author} {\bibfnamefont {P.}~\bibnamefont {Huillery}}, \bibinfo {author}
  {\bibfnamefont {P.}~\bibnamefont {Pillet}}, \bibinfo {author} {\bibfnamefont
  {D.}~\bibnamefont {Ciampini}}, \bibinfo {author} {\bibfnamefont
  {E.}~\bibnamefont {Arimondo}}, \ and\ \bibinfo {author} {\bibfnamefont
  {O.}~\bibnamefont {Morsch}},\ }\href {\doibase
  10.1103/PhysRevLett.113.023006} {\bibfield  {journal} {\bibinfo  {journal}
  {Phys. Rev. Lett.}\ }\textbf {\bibinfo {volume} {113}},\ \bibinfo {pages}
  {023006} (\bibinfo {year} {2014})}\BibitemShut {NoStop}%
\bibitem [{\citenamefont {Urvoy}\ \emph {et~al.}(2015)\citenamefont {Urvoy},
  \citenamefont {Ripka}, \citenamefont {Lesanovsky}, \citenamefont {Booth},
  \citenamefont {Shaffer}, \citenamefont {Pfau},\ and\ \citenamefont
  {L\"ow}}]{urvoy2015}%
  \BibitemOpen
  \bibfield  {author} {\bibinfo {author} {\bibfnamefont {A.}~\bibnamefont
  {Urvoy}}, \bibinfo {author} {\bibfnamefont {F.}~\bibnamefont {Ripka}},
  \bibinfo {author} {\bibfnamefont {I.}~\bibnamefont {Lesanovsky}}, \bibinfo
  {author} {\bibfnamefont {D.}~\bibnamefont {Booth}}, \bibinfo {author}
  {\bibfnamefont {J.~P.}\ \bibnamefont {Shaffer}}, \bibinfo {author}
  {\bibfnamefont {T.}~\bibnamefont {Pfau}}, \ and\ \bibinfo {author}
  {\bibfnamefont {R.}~\bibnamefont {L\"ow}},\ }\href {\doibase
  10.1103/PhysRevLett.114.203002} {\bibfield  {journal} {\bibinfo  {journal}
  {Phys. Rev. Lett.}\ }\textbf {\bibinfo {volume} {114}},\ \bibinfo {pages}
  {203002} (\bibinfo {year} {2015})}\BibitemShut {NoStop}%
\bibitem [{\citenamefont {Jones}\ \emph {et~al.}(1999)\citenamefont {Jones},
  \citenamefont {Lett}, \citenamefont {Tiesinga},\ and\ \citenamefont
  {Julienne}}]{jones1999}%
  \BibitemOpen
  \bibfield  {author} {\bibinfo {author} {\bibfnamefont {K.~M.}\ \bibnamefont
  {Jones}}, \bibinfo {author} {\bibfnamefont {P.~D.}\ \bibnamefont {Lett}},
  \bibinfo {author} {\bibfnamefont {E.}~\bibnamefont {Tiesinga}}, \ and\
  \bibinfo {author} {\bibfnamefont {P.~S.}\ \bibnamefont {Julienne}},\ }\href
  {\doibase 10.1103/PhysRevA.61.012501} {\bibfield  {journal} {\bibinfo
  {journal} {Phys. Rev. A}\ }\textbf {\bibinfo {volume} {61}},\ \bibinfo
  {pages} {012501} (\bibinfo {year} {1999})}\BibitemShut {NoStop}%
\bibitem [{\citenamefont {Beterov}\ \emph {et~al.}(2009)\citenamefont
  {Beterov}, \citenamefont {Tretyakov}, \citenamefont {Ryabtsev}, \citenamefont
  {Entin}, \citenamefont {Ekers},\ and\ \citenamefont
  {Bezuglov}}]{beterov2009}%
  \BibitemOpen
  \bibfield  {author} {\bibinfo {author} {\bibfnamefont {I.~I.}\ \bibnamefont
  {Beterov}}, \bibinfo {author} {\bibfnamefont {D.~B.}\ \bibnamefont
  {Tretyakov}}, \bibinfo {author} {\bibfnamefont {I.~I.}\ \bibnamefont
  {Ryabtsev}}, \bibinfo {author} {\bibfnamefont {V.~M.}\ \bibnamefont {Entin}},
  \bibinfo {author} {\bibfnamefont {A.}~\bibnamefont {Ekers}}, \ and\ \bibinfo
  {author} {\bibfnamefont {N.~N.}\ \bibnamefont {Bezuglov}},\ }\href@noop {}
  {\bibfield  {journal} {\bibinfo  {journal} {New J. Phys.}\ }\textbf {\bibinfo
  {volume} {11}},\ \bibinfo {pages} {013052} (\bibinfo {year}
  {2009})}\BibitemShut {NoStop}%
\bibitem [{\citenamefont {Viteau}\ \emph
  {et~al.}(2008{\natexlab{b}})\citenamefont {Viteau}, \citenamefont {Chotia},
  \citenamefont {Comparat}, \citenamefont {Tate}, \citenamefont {Gallagher},\
  and\ \citenamefont {Pillet}}]{viteau2008b}%
  \BibitemOpen
  \bibfield  {author} {\bibinfo {author} {\bibfnamefont {M.}~\bibnamefont
  {Viteau}}, \bibinfo {author} {\bibfnamefont {A.}~\bibnamefont {Chotia}},
  \bibinfo {author} {\bibfnamefont {D.}~\bibnamefont {Comparat}}, \bibinfo
  {author} {\bibfnamefont {D.~A.}\ \bibnamefont {Tate}}, \bibinfo {author}
  {\bibfnamefont {T.~F.}\ \bibnamefont {Gallagher}}, \ and\ \bibinfo {author}
  {\bibfnamefont {P.}~\bibnamefont {Pillet}},\ }\href {\doibase
  10.1103/PhysRevA.78.040704} {\bibfield  {journal} {\bibinfo  {journal} {Phys.
  Rev. A}\ }\textbf {\bibinfo {volume} {78}},\ \bibinfo {pages} {040704}
  (\bibinfo {year} {2008}{\natexlab{b}})}\BibitemShut {NoStop}%
\bibitem [{\citenamefont {Cederbaum}\ \emph {et~al.}(1997)\citenamefont
  {Cederbaum}, \citenamefont {Zobeley},\ and\ \citenamefont
  {Tarantelli}}]{cederbaum1997}%
  \BibitemOpen
  \bibfield  {author} {\bibinfo {author} {\bibfnamefont {L.~S.}\ \bibnamefont
  {Cederbaum}}, \bibinfo {author} {\bibfnamefont {J.}~\bibnamefont {Zobeley}},
  \ and\ \bibinfo {author} {\bibfnamefont {F.}~\bibnamefont {Tarantelli}},\
  }\href {\doibase 10.1103/PhysRevLett.79.4778} {\bibfield  {journal} {\bibinfo
   {journal} {Phys. Rev. Lett.}\ }\textbf {\bibinfo {volume} {79}},\ \bibinfo
  {pages} {4778} (\bibinfo {year} {1997})}\BibitemShut {NoStop}%
\bibitem [{\citenamefont {Amthor}\ \emph {et~al.}(2009)\citenamefont {Amthor},
  \citenamefont {Denskat}, \citenamefont {Giese}, \citenamefont {Bezuglov},
  \citenamefont {Ekers}, \citenamefont {Cederbaum},\ and\ \citenamefont
  {Weidem{\"u}ller}}]{amthor2009}%
  \BibitemOpen
  \bibfield  {author} {\bibinfo {author} {\bibfnamefont {T.}~\bibnamefont
  {Amthor}}, \bibinfo {author} {\bibfnamefont {J.}~\bibnamefont {Denskat}},
  \bibinfo {author} {\bibfnamefont {C.}~\bibnamefont {Giese}}, \bibinfo
  {author} {\bibfnamefont {N.~N.}\ \bibnamefont {Bezuglov}}, \bibinfo {author}
  {\bibfnamefont {A.}~\bibnamefont {Ekers}}, \bibinfo {author} {\bibfnamefont
  {L.~S.}\ \bibnamefont {Cederbaum}}, \ and\ \bibinfo {author} {\bibfnamefont
  {M.}~\bibnamefont {Weidem{\"u}ller}},\ }\href@noop {} {\bibfield  {journal}
  {\bibinfo  {journal} {Eur. Phys. J. D}\ }\textbf {\bibinfo {volume} {53}},\
  \bibinfo {pages} {329} (\bibinfo {year} {2009})}\BibitemShut {NoStop}%
\bibitem [{\citenamefont {B{\'e}guin}\ \emph {et~al.}(2013)\citenamefont
  {B{\'e}guin}, \citenamefont {Vernier}, \citenamefont {Chicireanu},
  \citenamefont {Lahaye},\ and\ \citenamefont {Browaeys}}]{beguin2013}%
  \BibitemOpen
  \bibfield  {author} {\bibinfo {author} {\bibfnamefont {L.}~\bibnamefont
  {B{\'e}guin}}, \bibinfo {author} {\bibfnamefont {A.}~\bibnamefont {Vernier}},
  \bibinfo {author} {\bibfnamefont {R.}~\bibnamefont {Chicireanu}}, \bibinfo
  {author} {\bibfnamefont {T.}~\bibnamefont {Lahaye}}, \ and\ \bibinfo {author}
  {\bibfnamefont {A.}~\bibnamefont {Browaeys}},\ }\href@noop {} {\bibfield
  {journal} {\bibinfo  {journal} {Phys. Rev. Lett.}\ }\textbf {\bibinfo
  {volume} {110}},\ \bibinfo {pages} {263201} (\bibinfo {year}
  {2013})}\BibitemShut {NoStop}%
\bibitem [{\citenamefont {Hankin}\ \emph {et~al.}(2014)\citenamefont {Hankin},
  \citenamefont {Jau}, \citenamefont {Parazzoli}, \citenamefont {Chou},
  \citenamefont {Armstrong}, \citenamefont {Landahl},\ and\ \citenamefont
  {Biedermann}}]{hankin2014}%
  \BibitemOpen
  \bibfield  {author} {\bibinfo {author} {\bibfnamefont {A.~M.}\ \bibnamefont
  {Hankin}}, \bibinfo {author} {\bibfnamefont {Y.-Y.}\ \bibnamefont {Jau}},
  \bibinfo {author} {\bibfnamefont {L.~P.}\ \bibnamefont {Parazzoli}}, \bibinfo
  {author} {\bibfnamefont {C.~W.}\ \bibnamefont {Chou}}, \bibinfo {author}
  {\bibfnamefont {D.~J.}\ \bibnamefont {Armstrong}}, \bibinfo {author}
  {\bibfnamefont {A.~J.}\ \bibnamefont {Landahl}}, \ and\ \bibinfo {author}
  {\bibfnamefont {G.~W.}\ \bibnamefont {Biedermann}},\ }\href@noop {}
  {\bibfield  {journal} {\bibinfo  {journal} {Phys. Rev. A}\ }\textbf {\bibinfo
  {volume} {89}},\ \bibinfo {pages} {033416} (\bibinfo {year}
  {2014})}\BibitemShut {NoStop}%
\bibitem [{\citenamefont {Schau\ss{}}\ \emph {et~al.}(2012)\citenamefont
  {Schau\ss{}}, \citenamefont {Cheneau}, \citenamefont {Endres}, \citenamefont
  {Fukuhara}, \citenamefont {Hild}, \citenamefont {Omran}, \citenamefont
  {Pohl}, \citenamefont {Gross}, \citenamefont {Kuhr},\ and\ \citenamefont
  {Bloch}}]{schaus2012}%
  \BibitemOpen
  \bibfield  {author} {\bibinfo {author} {\bibfnamefont {P.}~\bibnamefont
  {Schau\ss{}}}, \bibinfo {author} {\bibfnamefont {M.}~\bibnamefont {Cheneau}},
  \bibinfo {author} {\bibfnamefont {M.}~\bibnamefont {Endres}}, \bibinfo
  {author} {\bibfnamefont {T.}~\bibnamefont {Fukuhara}}, \bibinfo {author}
  {\bibfnamefont {S.}~\bibnamefont {Hild}}, \bibinfo {author} {\bibfnamefont
  {A.}~\bibnamefont {Omran}}, \bibinfo {author} {\bibfnamefont
  {T.}~\bibnamefont {Pohl}}, \bibinfo {author} {\bibfnamefont {C.}~\bibnamefont
  {Gross}}, \bibinfo {author} {\bibfnamefont {S.}~\bibnamefont {Kuhr}}, \ and\
  \bibinfo {author} {\bibfnamefont {I.}~\bibnamefont {Bloch}},\ }\href@noop {}
  {\bibfield  {journal} {\bibinfo  {journal} {Nature}\ }\textbf {\bibinfo
  {volume} {491}},\ \bibinfo {pages} {87} (\bibinfo {year} {2012})}\BibitemShut
  {NoStop}%
\bibitem [{\citenamefont {Greene}\ \emph {et~al.}(2000)\citenamefont {Greene},
  \citenamefont {Dickinson},\ and\ \citenamefont {Sadeghpour}}]{Greene2000}%
  \BibitemOpen
  \bibfield  {author} {\bibinfo {author} {\bibfnamefont {C.~H.}\ \bibnamefont
  {Greene}}, \bibinfo {author} {\bibfnamefont {A.~S.}\ \bibnamefont
  {Dickinson}}, \ and\ \bibinfo {author} {\bibfnamefont {H.~R.}\ \bibnamefont
  {Sadeghpour}},\ }\href@noop {} {\bibfield  {journal} {\bibinfo  {journal}
  {Phys. Rev. Lett.}\ }\textbf {\bibinfo {volume} {85}},\ \bibinfo {pages}
  {2458} (\bibinfo {year} {2000})}\BibitemShut {NoStop}%
\bibitem [{\citenamefont {Bendkowsky}\ \emph {et~al.}(2009)\citenamefont
  {Bendkowsky}, \citenamefont {Butscher}, \citenamefont {Nipper}, \citenamefont
  {Shaffer}, \citenamefont {L\"{o}w},\ and\ \citenamefont
  {Pfau}}]{Bendkowsky2009}%
  \BibitemOpen
  \bibfield  {author} {\bibinfo {author} {\bibfnamefont {V.}~\bibnamefont
  {Bendkowsky}}, \bibinfo {author} {\bibfnamefont {B.}~\bibnamefont
  {Butscher}}, \bibinfo {author} {\bibfnamefont {J.}~\bibnamefont {Nipper}},
  \bibinfo {author} {\bibfnamefont {J.~P.}\ \bibnamefont {Shaffer}}, \bibinfo
  {author} {\bibfnamefont {R.}~\bibnamefont {L\"{o}w}}, \ and\ \bibinfo
  {author} {\bibfnamefont {T.}~\bibnamefont {Pfau}},\ }\href@noop {} {\bibfield
   {journal} {\bibinfo  {journal} {Nature}\ }\textbf {\bibinfo {volume}
  {458}},\ \bibinfo {pages} {1005} (\bibinfo {year} {2009})}\BibitemShut
  {NoStop}%
\end{thebibliography}

%

\end{document}